\documentclass[prb,floats,superscriptaddress,showpacs]{revtex4}
\usepackage{amsmath}
\usepackage{graphicx}
\usepackage{setspace}

\renewcommand{\ss}{\mbox{\boldmath $\sigma$}}
\newcommand{\bfxi}{\mbox{\boldmath $\xi$}}
\newcommand{\bfnabla}{\mbox{\boldmath $\nabla$}}
\newcommand{\bfW}{{\bf W}}
\newcommand{\xx}{{\bf x}}
\newcommand{\xxp}{{\bf r}}
\newcommand{\for}{{\bf f}}
\newcommand{\vv}{{\bf v}}
\newcommand{\vvp}{{\bf u}}

\newcommand{\XX}{{\bf X}}
\newcommand{\VV}{{\bf V}}
\newcommand{\WW}{{\bf V'}}

\newcommand{\funo} {f^{(1)}}

\newcommand{\fudue} {f^{(2)}}
\newcommand{\fuenne} {f^{(N)}}

\newcommand{\fuesse} {f^{(s)}}
\newcommand{\bnabxn}{{\bfnabla_{{\bf r}_n}}}
\newcommand{\bnabvn}{{\bfnabla_{{\bf u}_n}}}
\newcommand{\bnabxin}{{\bfnabla_{{\bf x}_n}}}
\newcommand{\bnabvin}{{\bfnabla_{{\bf v}_n}}}

\newcommand{\bnabxx}{{\bfnabla_{{\bf X}}}}
\newcommand{\bnabvv}{{\bfnabla_{{\bf V}}}}
\newcommand{\partxi}{\partial_{X_i}}
\newcommand{\partxj}{\partial_{X_j}}
\newcommand{\partxk}{\partial_{X_k}}

\newcommand{\shat} {\hat{\ss}}
\newcommand{\CC}{{\bf{Y}}}
\newcommand{\GG}{{\bf{Z}}}
\newcommand{\bea}{\begin{eqnarray}}
\newcommand{\eea}{\end{eqnarray}}
\newcommand{\beq}{\begin{equation}}
\newcommand{\eeq}{\end{equation}}

\begin{document}
\date{\today}


\title{Phase-space approach to dynamical density functional theory}

\date{\today}

\author{Umberto Marini Bettolo Marconi}

\address{Dipartimento di Fisica, Universit\`a di Camerino and Istituto
Nazionale di Fisica della Materia, Via Madonna delle Carceri, 62032 ,
Camerino, Italy }
\author{ Simone Melchionna} 
\address{INFM-SOFT, Dipartimento di Fisica,  Universit\`a di Roma and Istituto
Nazionale di Fisica della Materia,  Piazzale A. Moro 2, 00185,
Roma, Italy }
 
\date{\today}
\begin{abstract}
We consider a system of interacting particles subjected to Langevin 
inertial dynamics 
and derive the governing time-dependent equation for the one-body density.
We show that, after suitable truncations of the 
Bogoliubov-Born-Green-Kirkwood-Yvon 
hierarchy, and a multiple time scale analysis,
we obtain a self-consistent equation involving only the one-body density.
This study extends to arbitrary dimensions previous work on a one-dimensional
fluid and highlights the subtelties of kinetic theory in the derivation of
dynamical density functional theory.
\end{abstract}

\maketitle

\section{Introduction}

Equilibrium density functional theory (DFT) provides a very powerful
tool to investigate the structure of inhomogeneous 
fluids~\cite{Evans1,Evans2,HansenMcDonald}. For systems
at thermodynamic equilibrium, all the statistical properties can be
expressed in terms of the average density field, $\rho(\xx)$.  In
turn, $\rho(\xx)$ can be obtained by minimizing the grand potential
functional, $\Omega[\rho(\xx)]$, which for a given interaction
potential and external one-body potential, is unique. Although
$\Omega[\rho(\xx)]$ is known exactly only in few particular cases,
fairly good approximations can be devised, so that the method is
versatile and generally applicable with success.

In out of equilibrium situations, there is no rigorous analogue of the free
energy, so that the description must be necessarily more complex.
In many instances, the density field is not sufficient to represent the state
of a system, and one has to identify the minimal set of relevant
fields which provide a satisfactory description of its evolution.  In
the past, phenomenological equations have been devised to deal with a
large class of non equilibrium processes, such as Navier-Stokes,
reaction-diffusion, Ginzburg-Landau equations, etc~\cite{Reichl, 
Kreuzer,DeGroot}.  
Standard
hydrodynamics, for instance, identifies $d+2$ relevant fields, number, momentum
and energy density, where $d$ is the spatial dimensionality of the
embedding space, each of these fields satisfying a conservation
law, and thus relaxing slowly toward equilibrium.  
On the other hand,
kinetic theory provides a truly microscopic derivation
of the Navier-Stokes equations and requires some assumptions about
the evolution of the phase space distribution~\cite{McLennan}.

When the dynamics of the particles is dissipative and overdamped, as
in the case of colloids strongly coupled to the solvent, one may assume
that the evolution of the density field is sufficient to capture the
relevant physics of the system.  This occurs because the momentum and
energy fluxes are not conserved  any more and thus become rapidly
enslaved by the density field.  If the friction coefficient, $\gamma$, 
is large, the
particles velocities rapidly relax in a  time of order
$\gamma^{-1}$, so that the evolution on longer timescales is only
determined by the dynamics of the configurational contribution 
to the distribution.
This is the point of view of the Smoluchowski description 
\cite{Smoluchowski,Gardiner,Risken}, which takes
into account only the configurational degrees of freedom.  Based on
these considerations, in recent years a dynamical density functional
theory (DDFT) has been put forward by several 
authors~\cite{Munakata}-\cite{Rex2006}.  In DDFT the
evolution of the noise-averaged, time-dependent, one-particle density
is completely deterministic and is driven solely by the gradient of
the equilibrium free energy functional.

In more general terms, inertial effects can play a role in a
number of different physical systems, as in the case of colloidal
particles weakly coupled to the solvent. Inertial effect must be
accounted for by means of the Kramers description in full phase space
\cite{Kramers,VanKampen}.
In the present paper, we derive a deterministic evolution equation for
the density field starting from the Kramers equation.  A second goal of the
present derivation is to show that the kinetic approach leads quite
naturally to an evolution equation which contains DDFT as a limiting
case.

There are a few reasons for working in the framework of the Kramers
equation.  The first is that Kramers is a more general description
since it contains as limiting cases the Liouville
equation and the fully overdamped dynamics.  
In addition, the Kramers equation
allows to deal quite naturally with systems under temperature
gradients \cite{Lopez}.
 Finally, from a fundamental
point of view, it seems to us rewarding to reconcile the density functional
approach, whose major success has been in explaining equilibrium
phenomena, with the kinetic approach which is usually applied to
non equilibrium situations.
Whereas  the current treatments show that it is quite natural
to obtain the DDFT equations for overdamped dynamics, it is not straightforward
to derive the DDFT from the Liouville equation, such as
in the case of a dense liquid, where the
hydrodynamic currents carried by particles are not strongly damped by the
viscosity of the surrounding fluid.  
A previous work by Marconi and Tarazona \cite{Marconi2006} has shown that in
the case of  one dimensional hard-rods the full inertial dynamics gives
rise to DDFT in a perturbative analysis truncated at lower order.
However, the extension to $d$-dimensions of this  procedure is not trivial,
because the number of independent tensorial components of the
moments of the distribution  increases with the order, $n$, as $d^n$.

We briefly recall the
several approaches appeared in recent literature to build
the equation of evolution of the one-body density profile in a classical
fluid. 
One class of approaches 
starts from the fundamental Hamiltonian description
of the fluid and aims to derive an equivalent evolution equation
for the density~\cite{Archer2006}. By using the
Kawasaki-Gunton projection operator formalism~\cite{KawasakiGunton},
Yoshimori~\cite{Yoshimori} has recently derived 
an equation of motion for the density
which differs from standard DDFT and is stochastic on account of the
coupling with fast degrees of freedom not explicitly retained.
Another class of equations for the density is obtained 
starting from the Langevin microscopic dynamics of the particles.
Within this class, some authors have kept the stochastic character
of the evolution of the instantaneous density~\cite{Dean}-\cite{Kawasaki2006}, 
whereas others
have focused attention on an ensemble averaged density and obtained
a purely deterministic equation. Both methods
do not require spatio-temporal coarse graining~\cite{ArcherRauscher}.

The different approaches underline the need for a theoretical
framework which is capable of deriving equations similar to the DDFT
in systems described by second order thermostatted inertial dynamics, 
instead of the Langevin overdamped dynamics. 
To this purpose one must consider a phase space rather than
a configurational space description.
The natural framework turns out to be the kinetic theory which has
predated the modern liquid state approaches. Unfortunately, a Boltzmann 
equation suitably modified to include
the coupling to the heat bath
only applies to a dilute gas regime.
However, several authors~\cite{Beijeren, Brey, Lutsko} have extended
the kinetic approach to the dense regime, where
the harsh repulsion between particles is taken into account via 
a revised version of the
Enskog theory (RET).  Ernst and van Beijeren  
have obtained the RET equation for the distribution
function from the full phase space 
representation of the N-particle dynamics.
In addition, Ernst and van Noije have considered
the evolution of thermostatted systems~\cite{Ernst}. We assume this theory as
the starting point to derive the time evolution equation for the
one body density for hard-sphere systems.

The present paper is organized as follows: in section \ref{Model}
we describe the microscopic model and write the evolution
equation for the N-particle distribution function.
Subsequently, we contract the description to the one-particle
distribution function, using the BBGKY hierarchy~\cite{Kreuzer},
and separate the velocity from the spatial
dependence of the distribution obtaining an open hierarchy
for the moments.
In section \ref{explicit} we close the hierarchy of the distribution
functions in the case of two explicit models for the intermolecular
forces. In the case of soft potentials we apply a Random Phase
Approximation, whereas in the case of hard-core potentials 
we consider the Revised Enskog Approximation~\cite{Beijeren,Enskog}.
In section \ref{multiscale}
we apply the multi-scale method to close the open hierarchy of the
moments and to obtain an equation for the density of the system.

\section{Model}
\label{Model}

Let us consider an assembly of $N$ heavy particles 
suspended in a solution of light
particles and moving in a d-dimensional 
region with positions $\xxp_n$ and velocities $\vvp_n$ with $n=1,N$. 
Due to their small mass, the solvent particles perform rapid
motions so that their influence on the heavy particles
can be described according to Langevin's idea by means of 
an effective stochastic force. As a
result of such elimination of microscopic degrees of freedom
the heavy particles experience a viscous drag force proportional to 
their velocity plus a random stochastic acceleration.
In addition the $N$ heavy particles of identical mass $m$
may move under the action of an external force $\for_e(\xx)$ and  
interact through a pair potential  $U(|\xxp-\xxp'|)$, which for simplicity
we take as a continuous function~\cite{Cecconi}-\cite{Pagnani}.
The equations of motions are:
\begin{eqnarray}
\label{kramers-a}
&d\xxp_n &= \vvp_n dt \\
& m d \vvp_n & = \left[ 
\for_e(\xxp_n) 
- \sum_{m(\neq n)} \bnabxn U(|\xxp_n-\xxp_m|) 
- m \gamma \vvp_n 
\right]dt  + d\bfW_n(t)
\label{kramers-b}
\end{eqnarray}
where $d\bfW_n(t) = \bfxi_n(t) dt$ is the increment of the Wiener process and
$\bfxi_n(t)$ is a Gaussian white noise with properties
\bea
\langle \xi^i_{n}(t) \rangle  &=& 0 \nonumber \\
\langle \xi^i_{n}(t)
\xi^j_{m}(s) \rangle  &=& 2 \gamma m k_B T\delta_{mn}
\delta^{ij} \delta(t-s)
\eea
where $T$ is the ``heat-bath temperature'' 
and $\langle \cdot \rangle$ indicates 
the average over a statistical ensemble of noise 
realizations~\cite{Cecconi}.
We include the friction exerted by the solvent via the coefficient $\gamma$. 

We now derive an equation for the singly conditioned probability distribution
$p(\{\xx,\vv\},t|\{\xx_0,\vv_0\},0)$ defined as the probability 
that the random variables $\{\xxp(t),\vvp(t)\}$ will lie between 
$\{\xx,\vv\}$ and $\{\xx+d\xx,\vv+d\vv\}$ given that
$\{\xxp(0),\vvp(0)\}=\{\xx_0,\vv_0\}$~\cite{Gardiner}.

Let us consider the infinitesimal time evolution
of an arbitrary function $G(\{\xxp(t),\vvp(t)\})$ 
of the trajectory of the system \cite{Gardiner}
\bea
d G(\{\xxp(t),\vvp(t)\}) =\sum_n \Bigl\{\Bigl[\vvp_n\cdot \bnabxn
+\Bigl( \frac{\for_e(\xxp_n)}{m}-
\frac{1}{m}\sum_{m(\neq n)}\bnabxn U(|\xxp_n-\xxp_m|)-\gamma \vvp_n
\Bigl)\cdot\bnabvn 
\nonumber \\
+\frac{\gamma k_B T}{m} \bnabvn \cdot \bnabvn
\Bigl]dt+
\frac{1}{m} 
d\bfW_n(t)
\cdot\bnabvn
\Bigl\}
G(\{\xxp(t),\vvp(t)\} ) 
\label{many}
\eea
The average over the realizations of such a function reads
\bea
&&\frac{d}{dt}\langle G(\{\xxp(t),\vvp(t)\} ) \rangle =\nonumber\\
&& \Bigl\langle \Bigl[\sum_n \Bigl\{ \vvp_n\cdot \bnabxn
+\Bigl( \frac{\for_e(\xxp_n)}{m}-
\frac{1}{m}\sum_{m(\neq n)}\bnabxn U(|\xxp_n-\xxp_m|)-\gamma \vvp_n
\Bigl)\cdot\bnabvn \nonumber \\
&+&
\frac{\gamma k_B T}{m} \bnabvn \cdot \bnabvn
\Bigl] G(\{\xxp(t),\vvp(t)\} ) \Bigr\rangle 
\label{many2}
\eea

Since the random variables $\{\xxp,\vvp\}$ have the conditional 
probability density
$p(\{\xx,\vv\},t|\{\xx_0,\vv_0\},0)$, we can equally write
\bea
\langle G(\{\xxp(t),\vvp(t)\}) \rangle = \int d^N\xx d^N\vv G(\{\xx,\vv\})
p(\{\xx,\vv\},t|\{\xx_0,\vv_0\},0)
\eea
and
\bea
\frac{d}{dt} \langle G(\{\xxp(t),\vvp(t)\} ) \rangle =
\int d^N\xx d^N\vv G(\{\xx,\vv\})
\frac{\partial}{\partial t} p(\{\xx,\vv\},t|\{\xx_0,\vv_0\},0)
\eea

Similarly, the r.h.s. of equation (\ref{many}) can be expressed as
\bea
\int d^N\xx d^N\vv \sum_n \Bigl\{ \Bigl[
\vv_n\cdot \bfnabla_{\xx_n}
+\Bigl( \frac{\for_e(\xx_n)}{m}-
\frac{1}{m}\sum_{m(\neq n)}\bfnabla_{\xx_n} U(|\xx_n-\xx_m|)-\gamma \vv_n
\Bigl)\cdot
\bfnabla_{\vv_n}
\nonumber \\
+\frac{\gamma k_B T}{m} \bfnabla_{\vv_n} \cdot \bfnabla_{\vv_n}
\Bigl] \Bigl\}
G(\{\xx,\vv\} ) 
p(\{\xx,\vv\},t|\{\xx_0,\vv_0\},0)
\label{many2}
\eea
We next integrate by parts and discard surface terms in equation 
(\ref{many2}) and
use the arbitrariness of the function $G$. The result is
\bea
\Bigl(\frac{\partial}{\partial t}&+&\sum_n[ L_0^{(n)}-L^{(n)}_{FP}]
\Bigl) p(\{\xx,\vv\},t|\{\xx_0,\vv_0\},0) =
\nonumber \\
&& \frac{1}{m} \sum_n\sum_{m (\neq n)}\bnabxin 
U(|\xx_n-\xx_m|)\cdot\bnabvin p(\{\xx,\vv\},t|\{\xx_0,\vv_0\},0)
\label{many3}
\eea
where we have introduced the notation
\bea
L_0^{(n)}=\vv_{n}\cdot\bnabxin+\frac{\for_e(\xx_n)}{m}\cdot\bnabvin \\
L^{(n)}_{FP}=
\gamma \Bigr[\bnabvin\cdot \vv_n +\frac{k_B T}{m}\bnabvin\cdot\bnabvin \Bigr]
\eea

To make contact with the standard statistical mechanical approach,
we introduce the $\Gamma$ phase-space
distribution, $\fuenne(\{\xx,\vv\},t)$ as the
probability density of finding 
the many body system between
$\{\xx,\vv\}$ and $\{\xx+d\xx,\vv+d\vv\}$ at time $t$.
Knowing the value of $\fuenne$ at time $t=0$ we obtain its value at
time $t$ via the relation
\beq
\fuenne(\{\xx,\vv\},t) = \int d\xx_0 d\vv_0 p(\{\xx,\vv\},t|\{\xx_0,\vv_0\},0) 
\fuenne(\{\xx_0,\vv_0\},0)
\eeq
One concludes that  the
evolution equation for $\fuenne(\{\xx,\vv\},t)$  
has the same form as equation~(\ref{many3}) relative to 
$p(\{\xx,\vv\},t|\{\xx_0,\vv_0\},0)$, i.e.:
\bea
\Bigl(\frac{\partial}{\partial t}&+&\sum_n[ L_0^{(n)}-L^{(n)}_{FP}]
\Bigl) \fuenne(\{\xx,\vv\},t) =
\nonumber \\
&& \frac{1}{m} \sum_n\sum_{m (\neq n)}
\bnabxin U(|\xx_n-\xx_m|)
\cdot 
\bnabvin \fuenne(\{\xx,\vv\},t)
\label{many4}
\eea
If one sets $\gamma=0$, the term $L^{(n)}_{FP}$ drops and 
the dynamics turns out to be conservative and the evolution of 
$\fuenne$ is described by Liouville equation.

The description contained in $\fuenne$ is for any practical purpose
unnecessarily detailed, so that we are lead to consider
the 
reduced s-particle distribution functions
obtained by integrating over $2 d (N-s)$ degrees of freedom: 
\bea
\fuesse(\xx_1,..&,&\xx_s,\vv_1,..,\vv_s,t)
=\frac{N!}{(N-s)!}\int \prod_{n=s+1}^N d\xx_n d\vv_n \times \nonumber\\
&& \fuenne (\xx_1,..,\xx_s,\xx_{s+1}..,\xx_N, 
\vv_1,..,\vv_s,\vv_{s+1}..,\vv_N,t).
\label{h1}
\eea


Thus integrating equation~(\ref{many4}) over (N-1) particle's coordinates and
velocities we obtain the exact   
evolution equation for the one particle distribution
\bea
\frac{\partial}{\partial t} \funo (\xx,\vv,t)
+ L_0 
\funo (\xx,\vv,t) =
L_{FP} \funo (\xx,\vv,t)+k(\xx,\vv,t) 
\label{fokker1}
\eea
where, the left hand side of equation~(\ref{fokker1}) 
contains the free streaming of
the particles, while  the right hand side describes the interactions with
the heat bath and those among the particles. The interaction term
$k(\xx,\vv,t)$, in the case of continuous potentials, reads
\beq
k(\xx,\vv,t)=
\frac{1}{m}{\bfnabla_{\bf v}} \int d\xx'\int d\vv' 
\fudue (\xx,\vv,\xx',\vv',t){\bf \bfnabla_{\bf x}}
U(|\xx-\xx'|) 
\label{Ke} 
\eeq
For particles with hard core interactions the term ${\bf \bfnabla_{\bf x}}
U(|\xx-\xx'|)$ is undefined and we have to modify the treatment as briefly
explained in section \ref{explicit}.
We remark that $\funo$ is the p.d.f. in the $2d$ dimensional $\mu$-space and
equation~(\ref{fokker1}), supplemented
by equation~(\ref{Ke}), represents the first equation of the
Bogoliubov-Born-Green-Kirkwood-Yvon (BBGKY) hierarchy~\cite{Kreuzer}, which
connects the evolution of the $n$-particle distribution function,
to the distribution function for  $(n+1)$ particles. 
Since $\funo(\xx,\vv,t)$ depends on the two-particle distribution 
$f^{(2)}(\xx,\vv,\xx',\vv',t)$, some approximate closure is required
to obtain a workable scheme. 

Before closing this section three remarks are in order:
i) the stationary solution $\fuenne$ of equation~(\ref{many4}) 
reduces to the equilibrium Gibbs (N,V,T)  phase-space distribution.
ii) if we take the limit $\gamma=0$, i.e. decouple the system from the 
heat bath, the dynamics becomes micro-canonical, i.e. at constant energy, 
and the evolution of $\fuenne$ is completely reversible.
iii) The number density $\rho(\xx,t)$ can be obtained 
from the ensemble average
of the microscopic operator $\sum_n\delta(\xx-\xxp_n(t))$, so that
$\rho(\xx,t)=\int d\vv\funo(\xx,\vv,t)$, the average particle current is
${\bf j}(\xx,t)=\int d\vv\funo(\xx,\vv,t)\vv$ and so on.


In the following we shall employ the non dimensional set of variables
which are obtained by measuring the velocities in units of the thermal
velocity  $v_T=\sqrt{k_B T/m}$ 
and lengths in unit of $\sigma$, i.e. $\VV\equiv\frac{\vv}{v_T}$ and 
$\XX\equiv\frac{\xx}{\sigma}$. 
The remaining variables can be non-dimensionalized
according to the transformations
$\tau\equiv t\frac{v_T}{\sigma}$, $\Gamma\equiv\gamma\frac{\sigma}{v_T}$,
${\bf F}(\XX)\equiv\frac{\sigma }{m v_T^2}{\bf f}_{e}(\xx)$.
Finally, the distribution function and the collision term are
rescaled according to the transformations:
$P(\XX,\VV,\tau)\equiv \sigma^d v_T^d \funo(\xx,\vv,t)$ and
$K(\XX,\VV,\tau)\equiv \sigma^{d+1} v_T^{d-1} k(\xx,\vv,t)$, where $d$ is the
dimensionality of the embedding space.


By rewriting equation (\ref{fokker1}) as 

\bea
\frac{1}{\Gamma}\frac{\partial P(\XX,\VV,\tau)}{\partial \tau}
&=&  {\cal L}_{FP} P(\XX,\VV,\tau)
-\frac{1}{\Gamma}\VV\cdot {\bf{\bfnabla_X}} P(\XX,\VV,\tau)\nonumber\\
&-&\frac{1}{\Gamma}{\bf F}(\XX,\tau)\cdot {\bf{\bfnabla_V}}
P(\XX,\VV,\tau)+\frac{1}{\Gamma} K(\XX,\VV,\tau)
\label{kramers0} 
\eea
one sees that, 
when the non dimensional damping constant $\Gamma$ is
large, the Fokker-Planck term 
\beq
{\cal L}_{FP} P(\XX,\VV,\tau) \equiv {\bf \bfnabla_V}\cdot\Bigl[
{\bf \bfnabla_V}+\VV\Bigl]  P(\XX,\VV,\tau)
\label{fokkerp}
\eeq
is dominant. This operator has non positive eigenvalues $\nu=0,-1,-2,..$
and corresponding eigenfunctions $H_{\underline{\alpha}}^{(\nu)}(\VV)$
defined as
\beq
 H_{\underline{\alpha}}^{(l)}(\VV)=h_{\underline{\alpha}}^{(l)}(\VV) \omega(V)
\label{hermite}
\eeq
where 
\beq
\omega(V)=\frac{1}{(2\pi)^{d/2}}e^{-\frac{1}{2}V^{2}}
\label{omega}
\eeq
and the subscript $\underline{\alpha}$ represents an array
of $\nu$ integer elements each ranging from $1$ to $d$~\cite{Harris,Moroni}.
The properties and the explicit representation 
of some of these polynomials are given in appendix A.

To separate the spatial dependence
from the velocity dependence
we expand the distribution function as : 



\bea
P(\xx,\vv,\tau) &=& \Phi^{(0)}(\xx,\tau)H^{(0)}(\vv)
+\Phi_{i}^{(1)}(\xx,\tau)  H_{i}^{(1)}(\vv) \nonumber \\
&+&\frac{1}{2!}\Phi_{ij}^{(2)}(\xx,\tau)H_{ij}^{(2)}(\vv)+\frac{1}{3!}
\Phi_{ijk}^{(3)}(\xx,\tau)H_{ijk}^{(3)}(\vv)+...
\eea
where the Einstein convention on repeated indices is used.
Similarly, we expand the interaction term and obtain
\beq
K(\xx,\vv,\tau)=C_{i}^{(1)}(\xx,\tau)H_{i}^{(1)}(\vv)
+\frac{1}{2!}C_{ij}^{(2)}(\xx,\tau)H_{ij}^{(2)}(\vv)
+\frac{1}{3!}C_{ijk}^{(3)}(\xx,\tau) H_{ijk}^{(3)}(\vv)+...
\eeq
The first term $C^{(0)}=0$ since the number of particles is conserved,
while the remaining are obtained from the formula:
\beq
C_{\underline{\alpha}}^{(l)}(\xx,\tau)\equiv \int d^{d}V 
h_{\underline{\alpha}}^{(l)}(\vv)K(\xx,\vv,\tau).
\label{collicoeff}
\eeq

Using the orthogonality of the tensorial Hermite polynomials 
one derives a set of differential equations for their coefficients
$\Phi_{\underline{\alpha}}^{(l)}(\XX,\tau)$.
This is achieved by multiplying both sides of equation (\ref{kramers0}) by 
$h_{\underline{\alpha}}^{(l)}$
and integrating with respect to $\VV$. One finds the following set of equations,
which reduce to the standard hydrodynamic ones for $\Gamma=0$:
\bea
\frac{\partial}{\partial\tau}\Phi^{(0)}(\XX,\tau) & = & 
-\partxi \Phi_{i}^{(1)}(\XX,\tau)
\label{grad1}
\eea
\bea
\frac{\partial}{\partial\tau}\Phi_{i}^{(1)}(\XX,\tau) & = &
-\Gamma \Phi_{i}^{(1)}(\XX,\tau)
-D_{i}\Phi^{(0)}(\XX,\tau)\nonumber\\
&&-\frac{1}{2}\partxj\Bigr(\Phi_{ji}^{(2)}(\XX,\tau)+\Phi_{ij}^{(2)}(\XX,\tau)
\Bigr)+C_{i}^{(1)}(\XX,\tau) 
\label{grad2}
\eea
\bea
\frac{\partial}{\partial\tau}\Phi_{ij}^{(2)}(\XX,\tau)
& =& -2\Gamma \Phi_{ij}^{(2)}(\XX,\tau)
-[D_{i}\Phi_{j}^{(1)}(\XX,\tau) + D_{j}\Phi_{i}^{(1)}(\XX,\tau)]\nonumber\\ 
& - & \partxk\Phi_{ijk}^{(3)}(\XX,\tau)+C_{ij}^{(2)}(\XX,\tau)
\label{grad3}
\eea
having introduced the short hand notation:
$$
D_{i}\equiv \partxi-F_i
$$

\section{Explicit models}
\label{explicit}

In order to proceed further we need to provide an
explicit representation of the collision
operator $K(\XX,\VV,\tau)$. We shall do that in two cases which may serve
to illustrate different aspects of the dynamics.

\subsection{Mean-field model}
Given a continuous and
differentiable pair potential $U(|\XX-\XX'|)$, we can write 
the collision kernel as
\beq
K(\XX,\VV,\tau)  =  \bnabvv \cdot \int d \VV' \int d \XX'
P^{(2)}(\XX,\XX',\VV,\WW,\tau)\bnabxx U(|\XX-\XX '|)
\eeq
which contains the  two particle distribution function 
$P^{(2)}(\XX,\XX',\VV,\WW,\tau) $.
We assume, as Enskog did,  that the doublet distribution function
$P^{(2)}(\XX,\XX',\VV,\WW,\tau)$
factors into the product:
$P^{(2)}(\XX,\XX',\VV,\WW,\tau)
=P(\XX,\VV,\tau)P(\XX',\WW,\tau)g_2(\XX,\XX'|\rho)$
in which the pair correlation $g_2$ is not a function of the velocities,
but only of the particle positions.
It is assumed that $g_2$ is a nonlocal equilibrium function of the profile
$\rho(\XX,t)$ and depends on time only through the density profile, and 
to have the same form as in a nonuniform equilibrium state 
whose density is $\rho(\XX,t)$.
Given the above, $g_2$ can be determined exactly from the second 
functional derivative of the 
free-energy functional for a inhomogeneous  system \cite{Evans2}.

With this approximation we find:
\beq
K(\XX,\VV,\tau) = -{\bf F}^{mol}(\XX,\tau)\cdot\bnabvv P(\XX,\VV,\tau)
\label{b2} 
\eeq
where, using $\rho(\XX,\tau)\sigma^d=\int d \VV' P(\XX',\WW,\tau)$,
we have introduced the molecular field
\bea
{\bf F}^{mol}(\XX,\tau) =
-\int d\XX' 
\rho(\XX',\tau)\
g_2(\XX,\XX'|\rho)
\bnabxx U(|\XX-\XX'|)   
\eea

In order to obtain the coefficients  $C_{\underline{\alpha}}^{(l)}$
one must take the scalar product 
of the operator  given by equation~(\ref{b2}) 
with $h_{\underline{\alpha}}^{(l)}(\VV)$. One finds: 
\bea
C^{(0)}(\XX,\tau) & = & 0\\
C^{(1)}_i(\XX,\tau) & = &-\Phi^{(0)} (\XX,\tau) F^{mol}_i(\XX,\tau)  
\label{vlasovcurrent}
\\
C^{(2)}_{ij}(\XX,\tau) & = & -\left[
\Phi_{i}^{(1)}(\XX,\tau)
F^{mol}_j(\XX,\tau) +\Phi_{j}^{(1)}(\XX,\tau)F^{mol}_i(\XX,\tau)\right] 
\eea

Let us show that at equilibrium the molecular field
can be expressed as a gradient. 
Following Marconi and Tarazona~\cite{Tarazona}, we consider the relation,
valid at thermodynamic equilibrium:
\bea
\int d\XX' \rho(\XX',\tau)g_2(\XX,\XX'|\rho) 
 \partxi U(\XX-\XX')
=\partxi
\frac{\delta {\cal F}^{exc}[\rho(\XX,\tau)]}
{\delta \rho(\XX,\tau)}
\nonumber \\
\label{b8} 
\eea
where ${\cal F}^{exc}$ is the free energy excess over the ideal gas,
so that the molecular field is of the form sought. Consequently, we express 
the first non vanishing component of the interaction as
\beq
C^{(1)}_i(\XX,\tau)
=\sigma^d
\rho(\XX,\tau)
\partxi\frac{\delta {\cal F}^{exc}[\rho(\XX,\tau)] }
{\delta \rho(\XX,\tau)}
\label{b9} 
\eeq

\subsection{Hard sphere model}
The hard sphere (HS) model can be viewed as a collection of hard,
perfectly elastic objects. There is no intermolecular force acting on them
except at the instant when two spheres are at distance $\sigma$, the HS 
diameter. Thus HS, due to the 
coupling with the stochastic heat bath, 
behave like non-interacting Brownian particles
between collisions. 
When two HS collide their velocities
after the impact, denoted with a prime, are related to those before
the impact (unprimed) velocities by
\bea
{\bf v}^{\prime}_1&=&{\bf v}_1-
({\bf v}_{12}\cdot\hat{\ss})\hat{\ss}
\nonumber \\
{\bf v}^{\prime}_2&=&{\bf v}_2+
({\bf v}_{12}\cdot\hat{\ss})\hat{\ss} 
\label{eq:uno}
\eea
where ${\bf v}_{12}={\bf v}_{1} -{\bf v}_{2}$, $\hat{\ss}$
is the unit vector directed from particle $1$ to particle $2$.

Using the fact that, although the derivative
of the intermolecular potential is ill-defined,
the trajectories in phase space are well defined,
Ernst and van Beijeren were able to derive the 
evolution equation
of the N-particle distribution function $\fuenne$~\cite{Ernst}. 
They proved that the evolution of the $\Gamma$ phase space 
distribution for a 
randomly driven HS fluid is governed  by the 
following pseudo-Liouville equation
\bea
\Bigl(\frac{\partial}{\partial t}+\sum_n L_0^{(n)}
\Bigl) \fuenne(\{\xx,\vv\},t)
=
\Bigl(\sum_n L^{(n)}_{FP}
+ 
\sum_{n<m}{\overline T}(nm)
\Bigr)
\fuenne(\{\xx,\vv\},t)
\nonumber \\
\label{dn}
\eea
The second term on the right hand side of equation~(\ref{dn}) 
contains the binary collision operator
${\overline T}(mn)$ for HS which accounts for the impulsive 
hard core interactions
and reads:
\beq
{\overline T}(mn)=\sigma^{d-1}
\int d\ss    
\Theta(\vv_{mn}\ss)(\vv_{mn}\cdot\ss)
\bigl[\delta(\xx_{mn}-\ss)b^{**}_{mn}-\delta(\xx_{mn}+\ss)\bigl]
\eeq
where $\Theta$ is the Heaviside function  
which selects only particles approaching
one another, 
$\vv_{mn}=(\vv_m-\vv_n)$, $\xx_{mn}=(\xx_m-\xx_n)$  and
$b^{**}_{mn}$ is the scattering operator defined for arbitrary function
$S(\vv_m,\vv_n)$ by 
\beq
b^{**}_{mn}S(\vv_m,\vv_n)=S(\vv_m',\vv_n')
\label{scat}
\eeq
which for elastic HS acts on the velocities $\vv_n$,
and replaces them by the restituting velocities as defined by 
equation~(\ref{eq:uno}).

Again starting from equation~(\ref{dn})
one obtains the first level of the BBGKY hierarchy
for the reduced one-particle distribution function $\funo(\xx,\vv,t)$
which is similar to equation~(\ref{fokker1}).
In the HS case
the derivation is quite lengthy
so that  we merely quote the final result of Ernst and van Beijeren 
for the interaction term
\bea
k(\xx_1,\vv_1,t)
&=& 
\sigma^{d-1}\int d\vv_2\int 
d\hat{\ss}\Theta(\hat{\ss}\cdot \vv_{12}) (\hat{\ss} 
\cdot \vv_{12})\times\nonumber\\
&&\{f^{(2)} (\xx_1,\vv_1',\xx_1-\ss,\vv_2',t)-
f^{(2)} (\xx_1,\vv_1,\xx_1+\ss,\vv_2,t)\}
\label{collision}
\eea
where the primes on the velocities denote scattered values determined 
from equation~(\ref{eq:uno}).
Finally, we consider the so-called Revised Enskog Theory (RET) 
which consists in replacing $f^{(2)}$,
using the {\em molecular chaos hypothesis}, 
\beq
f^{(2)}(\xx_1,\vv_1,\xx_2,\vv_2,t) \delta(|\xx_{12}|-\sigma)= 
\funo(\xx_1,\vv_1,t) \funo(\xx_2,\vv_2,t) g_2(\xx_1,\xx_2|\rho)
\delta(|\xx_{12}|-\sigma)
\nonumber \\
\label{factor}
\eeq
where $g_2(\xx_1,\xx_2|\rho)$ is the pair correlation function.
It is worth noticing the difference with the mean-field approximation.
Here, it is assumed that atoms are uncorrelated immmediately prior to
collision but are correlated after they collide \cite{Lutsko}, because collision itself 
generates correlations.

We turn hereafter to the nondimensional notation and use capital letters 
for the  reduced variables.
With the substitution (\ref{factor})
the collision integral $K(\XX,\VV,\tau)$ 
can be represented by the following
convolution product:
\bea
K(\XX_1,\VV_1,\tau)=
\int d\VV_2 \int
 d\hat{\ss}\Theta(\hat{\ss}\cdot \VV_{12})(\hat{\ss} \cdot \VV_{12})
\times
\nonumber\\
\Bigl\{g_{2}(\XX_1,\XX_1-\shat|\rho) 
P(\XX_1,\VV_1',\tau)P(\XX_1-\shat,\VV_2',\tau)-
\nonumber\\
g_{2}(\XX_1,\XX_1+\shat|\rho)P(\XX_1,\VV_1,\tau)
P(\XX_1+\shat,\VV_2,\tau)\Bigl\}
\eea


We now turn to the evaluation of the coefficients of the collision integral
via the Hermite expansion (\ref{collicoeff}). 
Since this expansion involves an infinite number of coefficients, we need to 
resort to a suitable truncation scheme.
A physically motivated prescription consists in choosing a restricted basis in
the complete Hilbert-space of the functions $h_{\underline{\alpha}}^{(l)}$. 
To this purpose
we choose the following $2+d$ functions
\bea
\Bigr \{ h^{(0)}(\VV),h_{i}^{(1)}(\VV),{\tilde h}^{(2)}(\VV)\Bigr\}
\eea
with $\tilde h^{(2)}(\VV)= \sum_i h_{ii}^{(2)}(\VV)$,
corresponding to the $2+d$ hydrodynamic modes (density, current and energy)
which in the absence of thermostat are conserved.


In order to project $K(\XX,\VV,\tau)$ on this basis
it is convenient to
replace $(\VV_1,\VV_2,\shat)\rightarrow(\VV_1',\VV_2',-\shat)$
so that the projection can be written as
\bea
&&C_{\underline{\alpha}}^{(l)}(\XX,\tau) = 
\int d\VV_1\int d\VV_2\int d\hat{\ss}
\Theta(\hat{\ss}\cdot \VV_{12})(\hat{\ss}\cdot \VV_{12})\times\nonumber\\
&&[h_{\underline{\alpha}}^{(l)}(\VV_1')-h_{\underline{\alpha}}^{(l)}(\VV_1)]
g_{2}(\XX,\XX+\shat|\rho)P(\XX,\VV_1,\tau)P(\XX+\shat,\VV_2,\tau)
\label{cgeneric}
\eea
Equation (\ref{cgeneric}) still represents a 
cumbersome $(3d-1)$-fold integral. However, by a change of variables 
and using simple properties of Gaussian integrals
it is possible to perform $2d$ integrations. In order to see that,
we  introduce
the relative velocity and the center of mass velocity:
\bea
\GG & = & \VV_1-\VV_2\nonumber\\
\CC & = & \frac{\VV_1+\VV_2}{2}
\eea
and notice that the Jacobian of transformation from $(\VV_1,\VV_2$) to $(\CC,\GG)$ 
is unity, so that the integral becomes,
using the Hermite representation of $P(\XX,\VV,\tau)$,
\bea
C^{(l)}_{\underline{\alpha}} (\XX,\tau)  =  
\int d\hat{\ss}
g_{2}(\XX,\XX+\shat|\rho)\int d^{d}\CC
\frac{1}{\pi^{d/2}}e^{-\CC^{2}}\int d^{d}
\GG\frac{1}{2^{d}
\pi^{d/2}}e^{-\frac{1}{4}\GG^{2}}\nonumber\\
\Theta(\hat{\ss}\cdot\GG)
(\hat{\ss}\cdot\GG) 
\left[h^{(l)}_{\underline{\alpha}}(\CC+\frac{\GG}{2}
-(\hat{\ss}\cdot\GG)\hat{\ss})-
h^{(l)}_{\underline{\alpha}}(\CC+\frac{\GG}{2}) \right]  \times\nonumber\\
\left[\Phi^{(0)}(\XX+\shat,\tau)+\sum_j h_{j}^{(1)}(\CC-\frac{\GG}{2})
\Phi_{j}^{(1)}(\XX+\shat,\tau)+
\frac{1}{2}\tilde h^{(2)}
(\CC-\frac{\GG}{2})\tilde \Phi^{(2)}(\XX+\shat,\tau)  \right]\times\nonumber\\
\left[\Phi^{(0)}(\XX,\tau)+\sum_k h_{k}^{(1)}(\CC+\frac{\GG}{2} ) 
\Phi_{k}^{(1)}(\XX,\tau)
+\frac{1}{2}\tilde h^{(2)}(\CC+\frac{\GG}{2})
\tilde \Phi^{(2)}(\XX,\tau) \right]
\eea
with $\tilde \Phi^{(2)}=\sum_i \Phi_{ii}^{(2)}$.
The integration over $\GG$ and $\CC$ can be performed straightforwardly
with the results:
\beq
C^{(0)}(\XX,\tau)=0
\eeq
\bea
&& C_i^{(1)}(\XX,\tau)= -\int d^{d}\hat{\sigma}
\hat{\sigma}_{i}g_{2}(\XX,\XX+\shat|\rho)
\Bigr\{\Phi^{(0)}(\XX,\tau)\Phi^{(0)}(\XX+\shat,\tau)+\nonumber\\
 &  & +\frac{2}{\sqrt{\pi}}\sum_{j=1}^{d}\hat{\sigma}_{j}
\left[
\Phi_{j}^{(1)}(\XX,\tau)\Phi^{(0)}(\XX+\shat,\tau)
-\Phi_{j}^{(1)}(\XX+\shat,\tau)\Phi^{(0)}(\XX,\tau)
\right]+\nonumber\\
&&\Phi^{(0)}(\XX,\tau)\tilde \Phi^{(2)}(\XX+\shat,\tau)+
\tilde \Phi^{(2)}(\XX,\tau)\Phi^{(0)}(\XX+\shat,\tau)+\nonumber\\
&&-\sum_{j,k=1}^{d}
\left[\frac{1}{2d}\delta_{jk}\Phi_{j}^{(1)}(\XX,\tau)\Phi_{k}^{(1)}(\XX+\shat,\tau)
-\frac{3}{2}\hat{\sigma}_{j}\hat{\sigma}_{k}
\Phi_{j}^{(1)}(\XX+\shat,\tau)\Phi^{(1)}_{k}(\XX,\tau)
\right]+\nonumber\\
&&+\frac{1}{\sqrt{\pi}}\sum_{j=1}^{d}\hat{\sigma}_{j}
\left[
\Phi_{j}^{(1)}(\XX,\tau)\tilde\Phi^{(2)}(\XX+\shat,\tau)
-\Phi_{j}^{(1)}(\XX+\shat,\tau)\tilde\Phi^{(2)}(\XX,\tau)
\right]
\Bigr\}
\eea

\bea
&& C^{(2)}(\XX,\tau)= -\int d^{d}\hat{\sigma}
g_{2}(\XX,\XX+\shat|\rho)\times\nonumber\\
&&\Bigr\{\sum_{j=1}^{d}\hat{\sigma}_{j}
\left[\Phi_{j}^{(1)}(\XX,\tau)\Phi^{(0)}(\XX+\shat,\tau)
+\Phi_{j}^{(1)}(\XX+\shat,\tau)\Phi^{(0)}(\XX,\tau)\right]\nonumber\\
&&\frac{4}{\sqrt{\pi}}\left[\Phi^{(0)}(\XX,\tau)\tilde \Phi^{(2)}(\XX+\shat,\tau)-
\tilde \Phi^{(2)}(\XX,\tau)\Phi^{(0)}(\XX+\shat,\tau)\right]+\nonumber\\
&&\sum_{j=1}^{d}\hat{\sigma}_{j}
\left[\Phi_{j}^{(1)}(\XX,\tau)\tilde\Phi^{(2)}(\XX+\shat,\tau)
+\Phi_{j}^{(1)}(\XX+\shat,\tau)\tilde\Phi^{(2)}(\XX,\tau)\right]
\Bigr\}
\eea
Having now obtained the expressions of the collision matrix elements
in the restricted hydrodynamic space, we could in principle 
insert these into the moment hierarchy equations (\ref{grad1}-\ref{grad3}). 
However, the hierarchy
is not closed and we need some prescription in order to 
truncate the expansion. 
In the following we shall consider a perturbative method to achieve this goal.


\section{Multiscale method}
\label{multiscale}
The multiple time-scale approach~\cite{Nayfeh}-\cite{Hansen}
represents a powerful perturbative
method to determine the temporal evolution
of the distribution function $P(\XX,\VV,\tau)$ in the high friction limit,
$\Gamma^{-1}<<1$.  
Physically,
such a limit means that many solute-solvent 
collisions take place while a solute 
particle travels a distance of the order of its radius.
The method is perturbative and the relevant expansion parameter
is the inverse dimensionless friction, $\Gamma^{-1}$.
By means of such a procedure Titulaer~\cite{Titulaer},
starting from the Kramers equation,
has derived the Smoluchowski equation for a colloidal particle.
The latter equation could have also be obtained, without 
resorting to the multiple time scale method, by simply taking
the high friction limit in the microscopic Langevin equation, i.e.
dropping the inertial term. However, the perturbative method
allows to compute the corrections to the Smoluchowski equation
when $\Gamma$ is finite~\cite{Wilemski}.

To begin with, the moments $\Phi_{\underline{\alpha}}^{(l)}$, 
their time derivatives
and the collision integral are expanded in powers of 
$\Gamma^{-1}<<1$:
\bea
\frac{\partial}{\partial\tau} & = &
\frac{\partial}{\partial\tau_{0}}
+\frac{1}{\Gamma}\frac{\partial}{\partial\tau_{1}}
+\frac{1}{\Gamma^{2}}\frac{\partial}{\partial\tau_{2}}+...
\label{m1}
\eea
\bea
\Phi_{\underline{\alpha}}^{(l)} & = &
\phi_{0,{\underline\alpha}}^{(l)}
+\frac{1}{\Gamma}\phi_{1,{\underline \alpha}}^{(l)}
+\frac{1}{\Gamma^{2}}\phi_{2,\underline{\alpha}}^{(l)}+...
\label{m2}
\eea
\bea
C_{\underline{\alpha}}^{(l)} & = &
c_{0,{\underline \alpha}}^{(l)}
+\frac{1}{\Gamma}c_{1,\underline{\alpha}}^{(l)}
+\frac{1}{\Gamma^{2}}c_{2,{\underline \alpha}}^{(l)}+...
\label{m3}
\eea


Let us remark that,
in order to construct the solution, we have replaced the single
physical time scale, $\tau$, by a set of auxiliary time scales
($\tau_0,\tau_1,..,\tau_n$) which are related to the original variable
by the relations $\tau_n=\Gamma^{-n}\tau$ and  
are treated as independent variables.
Also, all time-dependent functions of $\tau$ are  
replaced by  auxiliary functions of $(\tau_0,\tau_1,..)$.
Once the equations corresponding to the various orders have been 
determined, we return to the original time variable and functions.

Following the method of 
reference~\cite{Titulaer}, the amplitudes 
$\phi_{n}^{(0)}$ with $n>0$
are set
equal to zero.
Such a choice, although not unique, is sufficient to eliminate secular terms,
i.e. terms containing a dependence on the slow time $\tau_0$.

Expansions (\ref{m1}-\ref{m3}) must, now, be inserted into equation 
(\ref{kramers0}) 
in order to identify on both sides of the resulting equation
terms belonging to the same order in the 
expansion parameter $\Gamma^{-1}$ and in the Hermite polynomials.
We do not report the lengthy but straightforward derivation
but merely write down the 
set of equations which allows to relate all
the amplitudes  $\phi_{n,\underline{\alpha}}^{(l)}$ in terms
of the evolution of the amplitude $\phi_{0}^{(0)}$.

The order $\Gamma^{0}$ yields:
$\phi^{(0)}_0\not=0$ 
and $\phi^{(l)}_0=0~\forall l>0$. 

The first power $\Gamma^{-1}$ instead gives three equations,
which allow to determine the amplitudes of order $1$: 
\beq
\frac{\partial}{\partial\tau_{0}}\phi^{(0)}_0=0,
\eeq
 
\beq
-\phi^{(1)}_{1,i}=D_{i}\phi^{(0)}_{0}-c^{(1)}_{0,i}
\label{p11}
\eeq
and
\beq
\phi^{(2)}_{1,ij}=0
\eeq

The following order $\Gamma^{-2}$ gives and equation for the
partial $\tau_1$ derivative of the amplitude $\phi^{(0)}_0$ in terms
of a spatial derivative of the amplitude $\phi^{(1)}_{1,i}$: 
\beq
\frac{\partial}{\partial\tau_{1}}\phi^{(0)}_0
=-\partxi \phi^{(1)}_{1,i}
\eeq
Thus, using equation~(\ref{p11}), we obtain
\beq
\frac{\partial}{\partial\tau_{1}}\phi^{(0)}_0
=\partxi [D_{i}\phi^{(0)}_0-c_{0,i}^{(1)}]
\label{gam1}
\eeq

At the same order we also find the relation
\beq
\frac{\partial}{\partial\tau_{0}}\phi^{(1)}_{1,i}
=-\phi_{2,i}^{(1)}+c^{(1)}_{1,i}=0
\eeq
which using the fact that $\phi^{(1)}_{1,i}$ does not depend on
$\tau_0$, being a functional of $\phi^{(0)}_0$,
in view of equation~(\ref{p11}) leads to the equation:
\beq
\phi^{(1)}_{2,i}=c^{(1)}_{1,i}
\eeq
Finally, we find:
\beq
\phi_{\underline{\alpha}}^{1(l)}=0~\forall l>1
\eeq 

To order $\Gamma^{-3}$ we need to consider only the relation:
 
\beq
\frac{\partial}{\partial\tau_{2}}\phi^{(0)}_0
=-\partxi\phi^{(1)}_{2,i}=-\partxi c_{1,i}^{(1)}
\label{gam3}
\eeq
which allows to express the partial derivative  of  $\phi^{(0)}_0$
with respect to $\tau_2$ in terms of the collision integral.

We restore, now, the original time derivative: 
\begin{equation}
\frac{\partial}{\partial\tau}\phi^{(0)}_0=
(\frac{1}{\Gamma}
\frac{\partial}{\partial\tau_{1}}
+\frac{1}{\Gamma^{2}}\frac{\partial}{\partial\tau_{2}})
\phi^{(0)}_0
\end{equation}
and using equations~(\ref{gam1}) and (\ref{gam3}) we arrive at  
\begin{equation}
\frac{\partial}{\partial\tau}\phi^{(0)}_0(\XX,\tau)=
\frac{1}{\Gamma}\bnabxx\left[(\bnabxx-{\bf F}(\XX)) 
\phi^{(0)}_0(\XX,\tau)-{\bf c}_{0}^{(1)}(\XX,\tau)-\frac{1}{\Gamma}
{\bf c}_{1}^{(1)}(\XX,\tau)\right]
\label{final}
\end{equation}
 As one can see, at this
order in the perturbative expansion
only the component $\nu=1$ of the collision operator
appears in equation~(\ref{final}).
We need to give an explicit representation of $c_{n,i}^{(1)}$
in order to have a closed evolution equation.

In the Vlasov model the procedure is straightforward:
in fact, given eq. (\ref{vlasovcurrent}), the amplitudes $c_{n,i}^{(1)}$ only depend on the
$\phi^{(0)}_0$, i.e. the density profile.
Since such a quantity is of order $\Gamma^0$ in our 
small parameter expansion, it follows that there is no correction
of order $\Gamma^{-1}$. Only at order $\Gamma^{-2}$, which
is beyond the scope of the present treatment, one finds
a correction to the evolution equation in the Vlasov model.

In the hard-sphere model, instead, we have the two following
terms:
\begin{eqnarray}
&& c_{0,i}^{(1)}(\XX,\tau)= -\int d\hat{\ss}
\hat{\sigma}_{i}g_{2}(\XX,\XX+\shat|\phi_0^{(0)})
\phi_0^{(0)}(\XX,\tau)\phi_0^{(0)}(\XX+\shat,\tau)
\label{c01}
\end{eqnarray}

\begin{eqnarray}
c_{1,i}^{(1)}(\XX,\tau)&=& -\frac{2}{\sqrt{\pi}}\int d\hat{\ss}
\hat{\sigma}_{i}g_{2}(\XX,\XX+\shat|\phi_0^{(0)}) \times\nonumber\\
&&\sum_{j=1}^{d}\hat{\sigma}_{j}
[\phi_{1,j}^{(1)}(\XX,\tau)\phi_0^{(0)}(\XX+\shat,\tau)
-\phi_{1,j}^{(1)}(\XX+\shat,\tau)\phi_0^{(0)}(\XX,\tau)] \nonumber\\
\end{eqnarray}
with 
\beq
\phi_{1,i}^{(1)}(\XX,\tau)=-(\partxi-F_i(\XX))\phi_0^{(0)}(\XX,\tau)
+c_{0,i}^{(1)}(\XX,\tau).
\eeq

We show next that if we keep only the term $c_{0,i}^{(1)}$
and drop the term $c_{1,i}^{(1)}$ of order $\Gamma^{-1}$ in 
equation~(\ref{final})
we recover the time dependent
DDF equation. 
In fact, using the following expression
for the HS impulsive force:
$$
\partxi U_{HS}(|\XX-\XX'|)=-\hat{\sigma}_{i}\delta(|\XX-\XX'|-1),
$$ 
where $U_{HS}(|\XX-\XX'|)$ is the HS pair potential,
we can rewrite eq. (\ref{c01}) as 
\beq
c_{0,i}^{(1)}(\XX,\tau)
=\int d\XX'
g_{2}(\XX,\XX'|\phi_0^{(0)})
\phi_0^{(0)}(\XX,\tau)\phi_0^{(0)}(\XX',\tau)\partxi U_{HS}(|\XX-\XX'|)
\label{bnn} 
\eeq
or, using relation (\ref{b8}), in the equivalent form:
\beq
c_{0,i}^{(1)}(\XX,\tau)
=
\phi_0^{(0)}(\XX,\tau)
\partxi\frac{\delta {\cal F}_{HS}^{exc}[(\XX,\tau)] }
{\delta \phi_0^{(0)} (\XX,\tau)}
\label{b9} 
\eeq
where ${\cal F}_{HS}^{exc}$  is the excess contribution to the hard sphere free energy 
over the ideal gas value.
Finally, we remark that terms of higher 
order in $\Gamma^{-1}$ representing
dynamical corrections to the overdamped result, 
such as $c_{1,i}^{(1)}(\XX,\tau)$,
are functionals of $\phi_0^{(0)}(\XX,\tau)$, but cannot be expressed
as functional derivatives of the free energy. 
They play a role only when the system is out of equilibrium, i.e.
when currents are present. 

\section{Discussion}

In this paper we have derived the DDFT equation for the noise averaged
density $\rho(\xx,t)$ relative to a system of particles with inertia
and driven by a stochastic heat-bath. The starting point is the many
particle Kramers equation for the phase space distribution
which corresponds to the Langevin equations of motion.
Such a microscopic model describes fluids, where, due to the interactions
with the solvent, the momentum and energy reach equilibrium
on a faster scale than the slowly varying number density.
In more detail,
we employed the standard BBGKY hierarchy to derive an exact evolution
equation for the one particle distribution in terms of the two-particle
distribution function. 
In the case of  a soft intermolecular potential we have applied the so-called
Vlasov approximation to derive a closed equation for the distribution function.
On the other hand, in the case of hard spheres, in order 
to truncate the BBGKY hierarchy, we employed
the RET approximation, which is tantamount to factoring 
the two particle correlation into the product of two one-particle
phase space distributions times the equal time equilibrium configurational
two particle correlation function.
 The resulting equation is similar to the Enskog-Boltzmann equation 
for the one body phase-space distribution
augmented by a Fokker-Planck term accounting for the heat-bath.
 
In practical applications
a realistic intermolecular potential can always be approximated by a 
hard-sphere potential and a soft potential. 
Hence, by a slight modification of the theory presented here, 
one can treat the hard-core repulsion with the method illustrated
in section III.B, whereas the soft term is treated according to the method of
section III.A, assuming that the  processes induced by these 
forces proceed without interfering.

The further approximation of this paper is to use an
expansion in the inverse friction parameter to eliminate
the modes associated with momentum and energy fluxes. 
In order to avoid the appearance of secular terms this procedure is carried
out using a multiple-time scale analysis.

Interestingly, the method proposed yields,
to leading order in the inverse friction
parameter,  an equation identical to that obtained in the DDF theory. 
Moreover, the merit of the present approach is the possibility to describe
systems subject to externally imposed temperature gradients~\cite{Lopez} or
inelastic non conserving 
interactions~\cite{Marconi2006,Paolotti}. Finally, the theory allows to
compute higher order corrections, whose next order to DDF theory
represents the coupling between the density and the momentum current.
Given the above, our result could be employed to study the dynamics of 
systems where
inertial and structural effects result in a non trivial interplay, such as
for systems undergoing Poiseuille flow or under shear. Candidate systems are 
colloidal suspensions and dense confined fluids.

Finally, we wish to comment that the present theory can be useful for obtaining
information about the relaxation of the structural properties, but does
not allow to study hydrodynamic phenomena. In fact, Galilean invariance
and conservation of momentum are explicitly violated by our model.
The Langevin coupling dampens the absolute velocities due to 
the presence of a solvent at rest in the laboratory frame.
However, the conservation of momentum could be
enoforced by introducing a translationally invariant thermostat which
damps the relative instead of the absolute momenta of particles. 
Actually, this strategy is at the basis of the Dissipative Particle Dynamics
(DPD) ~\cite{Espanol} and could be embedded in the framework of the present
approach.











U.M.B.M. acknowledges the support of the 
Project COFIN-MIUR 2005, 2005027808.

\appendix

\section{Hermite polynomials}
The $\nu$-dimensional Hermite tensor polynomials, 
$h^{(\nu)}_{\underline{\alpha}}(\VV)$,
introduced in equation~(\ref{hermite}) are tensorial generalizations 
of the standard Hermite polynomials \cite{Arfken}.
These are defined through the formula
\begin{eqnarray*}
h^{(\nu)}_{\underline{\alpha}}(\VV)
\equiv(-1)^{\nu} e^{V^2/2}\partial_{V_{\alpha_{1}}}..
\partial_{V_{\alpha_{\nu}}}e^{-V^2/2}
\end{eqnarray*} 
Explicitly the first polynomials read
\begin{eqnarray*}
h^{(0)}(\VV) & = & 1\\
h_{i}^{(1)}(\VV) & = & V_{i}\\
h_{ij}^{(2)}(\VV)& = & (V_{i}V_{j}-\delta_{ij}) \\
h_{ijk}^{(3)}(\VV)& = &\left[(V_{i}V_{j}V_{k} -
(V_{i} \delta_{jk}+V_{j} \delta_{ik}+V_{k} \delta_{ij}) \right] \\
h_{ijkl}^{(4)}(\VV) & = &  [V_{i}V_{j}V_{k}V_{l}
-(V_{i}V_{j}\delta_{kl}+ V_{i}V_{k}\delta_{jl} 
+ V_{i}V_{l}\delta_{jk}+\\
&&V_{j}V_{k}\delta_{il}+V_{j}V_{l}\delta_{ik}
+V_{k}V_{l}\delta_{ij})+ (\delta_{ij}\delta_{kl}+
\delta_{ik}\delta_{jl}+\delta_{il}\delta_{jk})]  
\end{eqnarray*}

 The tensor
polynomials are orthonormal with respect to the weight
$\omega(V)$ defined by equation~(\ref{omega}),
\beq
\int d\VV\omega(V) h^{(\mu)}_{\underline{\alpha}}(\VV)
h^{(\nu)}_{\underline{\beta}}(\VV)=\delta_{\underline{\alpha}
\underline{\beta}}^{(\mu\nu)}
\eeq
where $\delta_{\underline{\alpha}\underline{\beta}}^{(\mu\nu)}$
is zero unless $\mu=\nu$ and the subscript $\underline{\alpha}$ 
is a permutation of the subscript $\underline{\beta}$.

\end{document}